\documentclass[preprint,showpacs,pre,amssymb]{revtex4-1}

\usepackage{graphicx}
\usepackage{amsmath}

\begin{document}

\title{Multicanonical distribution and the origin of power laws}

\author{G. L. Vasconcelos}
\email{giovani@df.ufpe.br}
\author{D. S. P. Salazar}
\email{dsps@df.ufpe.br}

\affiliation{Departamento de F\'{i}sica, Universidade Federal de Pernambuco, Brazil.}

\affiliation{Unidade de Educa\c{c}\~{a}o a Dist\^{a}ncia e Tecnologia, Universidade Federal Rural de Pernambuco, Brazil.}

\begin{abstract}
A multicanonical formalism is applied to the problem of statistical equilibrium in a complex system with a hierarchy of dynamical structures. At the small scales the system is in quasi-equilibrium and follows a Maxwell-Boltzmann distribution with a slowly fluctuating temperature. The probability distribution  for the temperature is determined using Bayesian analysis and it is then used to average the Maxwell-Boltzmann distribution.   The resulting energy distribution law is written in terms of generalized hypergeometric functions, which display power-law tails.
\end{abstract}


\maketitle

\section{INTRODUCTION}

Power law distributions occur in widely diverse physical 
systems spanning an impressive range of length scales \cite{review-power}.  
The appearance of heavy-tailed distributions is  often traced to the presence of hierarchical structures in the system 
\cite{pnas}, whose complex ``interaction'' may result in violation of the statistical independence of subsystems, thus leading to non-Gibbsian distributions \cite{treumann}. 
Despite these  insights and the many contributions to the problem \cite{li_site}, it is fair to say that the physical mechanisms behind the  emergence of power-law distributions are not yet well  understood. 

Recently, we introduced \cite{ourPRL2012} a  general formalism to describe statistical equilibrium of complex systems with multiple  scales where the probability distribution of states displays power-law tails. In this formalism, the hierarchical structure embedding the system of interest is effectively modeled as a set of nested ``internal heat reservoirs,'' where each ``reservoir" is described by only one effective degree of freedom, namely, its  ``temperature." On the basis of a few physically reasonable assumptions, it was possible to show that for a large class of systems the equilibrium distribution can be written explicitly in terms of certain generalized hypergeometric  functions, which exhibit power law tails. This family of generalized hypergeometric (GHG) distributions includes, as its first two members, the Boltzmann-Gibbs distribution and the Tsallis distribution \cite{tsallis1}.  Higher-order members of the GHG family of distributions have been shown to describe remarkably well the statistics of velocity fluctuations in  turbulence \cite{ourPRE2010}.

The GHG distribution represents a  generalization of the canonical distribution for multiscale systems and hence it is also called  multicanonical. The main purpose of  the present paper is to give an alternative derivation, based on Bayesian analysis, of the multicanonical distribution. 
Because the derivation of the multicanonical distribution given here  tries to parallel (whenever possible) the usual treatment of the canonical distribution, we shall begin our presentation by briefly reviewing the derivation of  the canonical distribution.
 
 \begin{figure}
\includegraphics[width=0.3\textwidth]{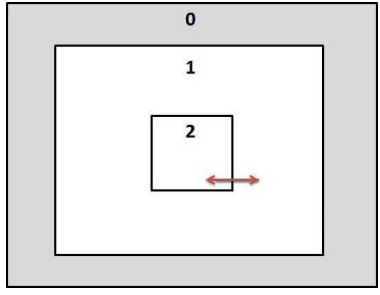}
\caption{\label{fig1} A canonical system in thermal equilibrium at temperature $T_0$. Arrows indicate energy exchange between the subsystems.}
\end{figure}

\section{The Canonical Distribution}

Consider  a  system  in thermal equilibrium at some temperature $T_0$.  We recall that in the canonical formalism the constant temperature constraint is enforced by embedding the system is a much larger system (i.e., a heat bath)  capable of giving it energy \cite{wannier}. We designate the system of interest by the label 2 and the larger system embedding it by the label 1, with  the combined system formed by subsystems 1 and 2 being given the label 0; see Fig.~1.

The energies of systems  1 and 2 will be denoted by $E_k^1$ and $E_i^2$, respectively, where $k$ and $i$ represent the labels designating the possible states in each system. Let us denote by $p_0(E_{k,i})=p_0(E_k^1+E_i^2)$ the probability of finding system 0 in a state corresponding to energy $E_{k,i}=E_k^1+E_i^2$. In view of the independence between systems 1 and 2, one can then write
\begin{align}
p_0(E_k+E_i)=p_1(E_k) p_2(E_i),
\label{eq:p0}
\end{align}
where $p_1(E_k)$ and $p_2(E_i)$ are the probabilities of finding systems 1 and 2 in states with energy $E_k$ and  $E_i$, respectively. Taking the logarithm derivative of (\ref{eq:p0}) with respect to $E_k$ yields
\begin{align}
 \frac{\partial \ln p_0(E_i+E_k)}{\partial E_k}=\frac{d \ln p_1(E_k)}{d E_k} \equiv -\beta_1,
 \label{eq:beta1}
 \end{align}
 where $\beta_1$ may be a  function of $E_k$ but {\it not} of $E_i$.
On the other hand, it is clear that 
\begin{align}
  \frac{\partial \ln p_0}{\partial E_k} = \frac{\partial \ln p_0}{\partial E_i}=  \frac{d \ln p_2(E_i)}{d E_i}.
  \label{eq:dpo}
 \end{align}
 Comparing (\ref{eq:beta1}) and (\ref{eq:dpo}), one then concludes that
\begin{align}
 \frac{d \ln p_2(E_i)}{d E_i}= -\beta_1.
\end{align}
Since $\beta_1$ does not depend on $E_i$, the preceding equation can be readily integrated, yielding
\begin{align}
 p_2(E_i|\beta_1) = \mbox{constant}\cdot \exp({-\beta_1 E_i}).
 \label{eq:p2}
\end{align}
Since the partition of system 0  into  subsystems 1 and 2 is entirely arbitrary, the quantity $\beta_1$ must be the same for any partition one chooses. In other words,  $\beta_1$ is a characteristic of system 0 only, that is,
\begin{align}
\beta_1=\beta_0,
\label{eq:b1}
\end{align}
which implies that 
\begin{align}
 p_2(E_i|\beta_0) = A \exp\left({-\beta_0E_i}\right),
 \label{eq:p1}
\end{align}
where $A$ is a constant.

It should be evident that the discussion above is completely symmetrical with respect to labels 1 and 2, so that system 2 (irrespective of its size) obeys the same distribution law as system 1. In then  follows from (\ref{eq:p0}) and (\ref{eq:p1}) that  the probability of finding any subsystem of system 0 in a state with energy $E$ is given by the Boltzmann-Gibbs (BG) distribution:
\begin{align}
p(E|\beta_0)=\frac{g(E)\exp(-{\beta_0E})}{Z_0(\beta_0)},
 \label{eq:BG}
\end{align}
where $g(E)$is the density of states  and 
\begin{equation}
Z_0(\beta)=\int_{0}^\infty g(E) \exp \left(-\beta{E}\right)dE
\label{eq:Z}
\end{equation}
is the  partition function

It is important to emphasize here that the key step in deriving (\ref{eq:BG}) was the ability to partition the system into two independent subsystems of arbitrary sizes, so that each subsystem is described by the same distribution law. There are however many physical systems, where the relevant probability distributions depend on the scale at which the measurements are made. In such complex systems, the system cannot be partitioned into independent subsystems of arbitrary sizes, and one has to treat  each dynamical scale separately, as discussed next.

\section{The Multicanonical Distribution}

Here we consider a multiscale system of size $L$ in thermal equilibrium at temperature $T_0$. We assume  that the system possesses a hierarchy of dynamical structures of characteristic sizes $\ell_i=L/b^{i-1}$, with $i=1,2,...,n$, where $b$ is a number greater than 1. (The specific value of $b$ is not relevant here.) 
We suppose furthermore that there is a wide separation of time scales within this hierarchy, with smaller structures having shorter characteristic times. 
Let us now consider a partition of our system into ``nested'' subsystems of sizes  $\ell_j$, 
as indicated in Fig.~2.    We shall designate the subsystem of size $\ell_j$ by the label $j$. 
The ``thermodynamic state'' of each subsystem $j$ will be characterized by only one parameter, namely,  its inverse temperature  $\beta_j$. As before, we designate  the combined system consisting of all subsystems by the label 0. 

\begin{figure}
\includegraphics[width=0.3\textwidth]{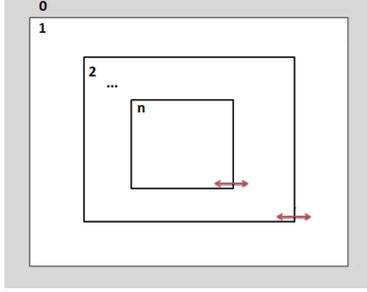}
\caption{\label{fig1} A multicanonical system in thermal equilibrium at temperature $T_0$. }
\end{figure}

Let us now focus our attention on the subsystem $n$ of size $\ell_n$. Since $\ell_n$ is the smallest characteristic length scale in the system, it is clear that this subsystem can be arbitrarily divided into two independent subsystems. Thus, repeating  the same reasoning that led to Eq.~(\ref{eq:BG}), one  obtains that the energy distribution law for  this subsystem  is  given by
\begin{align}
p(E|\beta_n)=\frac{g(E)\exp(-{\beta_nE})}{Z_0(\beta_n)}.
 \label{eq:BGn}
\end{align}
Note, however, that  owing  to the (intermittent) energy exchange between subsystem $n$ and its immediate surrounding, represented by subsystem $n-1$, the parameter $\beta_n$  is  no longer constant but rather will fluctuate  randomly.  If we denote by $f(\beta_n)$ the probability density function (PDF) of  $\beta_n$, then the marginal distribution $p(E)$ for subsystem $n$ reads
\begin{align}
 p(E)=g(E) \int_{0}^{\infty} \frac{\exp(-{\beta_nE})}{Z_0(\beta_n)}f(\beta_n)d\beta_n.
 \label{pE}
\end{align}

Next we wish to compute   $f(\beta_n)$.  To do so, we shall make use of Bayesian analysis \cite{gelman}. First recall that subsystem  $n$ is embedded in a much larger  subsystem ${n-1}$, characterized by the parameter $\beta_{n-1}$, which is assumed to vary much slower than $\beta_n$.  We are thus interested in computing $f(\beta_n|\beta_{n-1})$. From Bayes' theorem \cite{gelman} one has
\begin{equation}
f(\beta_n;\beta_{n-1}|E)\propto p(E|\beta_n)f(\beta_n;\beta_{n-1}),
\label{eq:f}
\end{equation}
where $\beta_{n-1}$ is considered a (hyper)parameter of the distribution of $\beta_n$.
In Bayesian parlance, the distribution $f(\beta_n;\beta_{n-1})$ is called the  prior distribution,  $p(E|\beta_n)$ is the likelihood function, and  $f(\beta_n;\beta_{n-1}|E)$ is the posterior distribution. 
[In Eq.~(\ref{eq:f}) we introduced the notation $f(\beta_n;\beta_{n-1})\equiv f(\beta_n|\beta_{n-1})$ for convenience.]

Let us assume, as is often done in Bayesian analysis,  that the prior distribution is conjugate to the  likelihood  $p(E|\beta_n)$, meaning that the  posterior distribution follows the same parametric form  as the prior distribution.  If we consider the rather general case where $g(E)\propto E^{\gamma-1}$, $\gamma>0$, so that $Z_0(\beta_n)\propto \beta_n^{-\gamma}$, it then follows from Eq.~(\ref{eq:BG}) 
that $p(E|\beta_n)$ is given by
\begin{equation}
p(E|\beta_n)\propto \beta_n^{\gamma}E^{\gamma-1}\exp\left(-\beta_n{E}\right).
\end{equation}
which when viewed as the likelihood of the parameter $\beta_n$ is proportional to a gamma distribution. Now, it is well known \cite{gelman} that in this case the conjugate prior is also a gamma distribution, and so we have\begin{equation}
\label{gamma}
f(\beta_n|\beta_{n-1})=\frac{1}{\beta_n\Gamma(\alpha+1)}\left(\frac{\alpha\beta_{n}}{\beta_{n-1}}\right)^{\alpha+1}\exp\left(-\frac{\alpha\beta_n}{\beta_{n-1}}\right),
\end{equation}
where $\alpha$ is a constant. In obtaining Eq.~(\ref{gamma}) we also used the fact that $\langle\beta_n|\beta_{n-1}\rangle=\beta_{n-1}$, as it should, since subsystem $n-1$ acts as a heat reservoir for subsystem $n$.
By scale invariance,  we assume that the distribution $f(\beta_j|\beta_{j-1})$, for $j=1,...,n$,   has the same form as in Eq.~(\ref{gamma}).  

We now have
\begin{equation}
f(\beta_n)=\int_0^\infty f(\beta_n|\beta_{n-1})f(\beta_{n-1})d\beta_{n-1}.
\end{equation}
Using this relation recursively then yields
\begin{equation}
\label{marginal}
f(\beta_n)=\int_{0}^{\infty}...\int_{0}^{\infty}\prod_{j=1}^{n}f(\beta_{j}|\beta_{j-1})\,d\beta_{1}\cdots d\beta_{n-1},
\end{equation}
with $f(\beta_j|\beta_{j-1})$  given by Eq.~(\ref{gamma}).
After inserting Eq.~(\ref{marginal})  into Eq.~(\ref{pE}), and performing a sequence of changes of variables of the type $x_j=\alpha\beta_{j}/\beta_{j-1}$,  one can show  that the resulting multidimensional integral can be expressed in terms of known higher transcendental functions:
 \begin{equation}
 p(E)  = \frac{g(E)}{Z_n} \,{ _{n}F_{0}}(\alpha+\gamma+1,...,\alpha+\gamma+1;-{\beta}_0\alpha^{-n} E),
 \label{eq:ghg}
\end{equation}
where $_{n}F_{0}(\alpha_{1}, ...,\alpha_{n}; -z)$ is the generalized hypergeometric function of order $(n,0)$ \cite{erdelyi}.
The small-scale partition function, $Z_n$, is given by
\begin{align}
Z_n   = Z_0(\beta_0)\left[ \frac{\alpha^\gamma\Gamma({\alpha+1})}{\Gamma(\alpha+\gamma+1)}\right]^n.
 \label{eq:A11}
\end{align}

One  important property of the generalized hypergeometric (GHG) distribution  given in (\ref{eq:ghg})  is that it exhibits power-law tails  of the form: $p(E)  \propto E^{-(\alpha+2)}$, for $E\to\infty$. This follows immediately from the asymptotic expansion   of the function ${_{n}F_{0}}$ \cite{wolfram}: ${_{n}F_{0}}(\alpha_{1},...,\alpha_{n};- x) = \sum_{i=1}^{n} C_{i}x^{-\alpha_n}\left(1+O(1/x)\right)$, as $x\to\infty$.  It is also worth pointing out that the first two members of the family $_{n}F_{0}$ yield elementary functions, namely,  $_{0}F_{0}(x)=\exp(x)$  and  $_{1}F_{0}(1/(q-1),x)=\exp_{q}\left(x/(q-1)\right)$,   where $\exp_{q}(x)$ is the $q$-exponential: $\exp_{q}(x)=[1+(1-q)x]^{1/(1-q)}$. The GHG distribution with $n=0$  thus recovers the Boltzmann-Gibbs distribution,  whereas for  $n=1$ it gives the  $q$-exponential or Tsallis distribution \cite{tsallis1}. One then sees from the preceding discussion that if a system with only one time scale  is in thermal equilibrium then the Boltzmann-Gibbs distribution follows, whereas if it has two distinct time scales the Tsallis distribution should be applicable.  For complex systems with more than two characteristic time scales, such as turbulent flows, GHG distributions of higher order are thus required \cite{ourPRE2010,ourPRL2012}.

\section{Conclusions}

We have presented an alternative derivation, based on Bayesian analysis, of the multicanonical distribution, which describes the statistical equilibrium of  complex systems possessing a hierarchy of time and length scales. We have shown that the multicanonical distribution  can be written explicitly in terms of generalized hypergeometric functions, which exhibit a power-law asymptotic behavior. This thus shows that the emergence of power law distributions---an ubiquitous feature in nature---is intimately connected with the existence of multiple  time and length scales in the system. 

\begin{acknowledgments}
This work was supported in part by the Brazilian agencies CNPq and FACEPE.
\end{acknowledgments}

{}

\end{document}